\documentclass{sig-alternate-05-2015}

\begin{document}

\setcopyright{acmcopyright}




%

\title{Role of Temporal Diversity in Inferring Social Ties Based on Spatio-Temporal Data}
%
%
%
%
%

\numberofauthors{3} 
%
\author{
%
%
\alignauthor
Deshana Desai\\
		\affaddr{Dhirubhai Ambani Institute of Information and Communication Technology, Gandhinagar}\\
       \affaddr{Gujarat, India}\\
       \email{desai.deshna@gmail.com}
\alignauthor
Harsh Nisar\\
       \affaddr{Institute of Financial Management and Research}\\
       \affaddr{New Delhi, India}\\
       \email{nisar.harsh@gmail.com}
\alignauthor
Rishab Bhardawaj\\
       \affaddr{Dhirubhai Ambani Institute of Information and Communication Technology, Gandhinagar}\\
       \affaddr{Gujarat, India}\\
       \email{bhardwaj.rish@gmail.com}
}

\maketitle
\begin{abstract}
The last two decades have seen a tremendous surge in research on social networks and their implications. The studies includes inferring social relationships, which in turn have been used for target advertising, recommendations, search customization etc. However, the offline experiences of human, the conversations with people and face-to-face interactions that govern our lives interactions have received lesser attention. We introduce \textbf{DAIICT Spatio-Temporal Network (DSSN), a spatiotemporal dataset} of 0.7 million data points of continuous location data logged at an interval of every 2 minutes by mobile phones of 46 subjects. 

Our research is focused at inferring relationship strength between students based on the spatiotemporal data and comparing the results with the self-reported data. In that pursuit we introduce \textbf{Temporal Diversity}, which we show to be superior in its contribution to predicting relationship strength than its counterparts. We also explore the evolving nature of Temporal Diversity with time. 

Our rich dataset opens various other avenues of research that require fine-grained location data with bounded movement of participants within a limited geographical area. The advantage of having a bounded geographical area such as a university campus is that it provides us with a microcosm of the real world, where each such geographic zone has an internal context and function and a high percentage of mobility is governed by schedules and time-tables. The bounded geographical region in addition to the age homogeneous population gives us a minute look into the active internal socialization of students in a university. 
\end{abstract}

\vfill

\section{Introduction}
The emergence of new technologies that enable collection of geospatial and temporal data along with the explosive growth of research in the fields of online social networks and their implications have led the path to conduct research on the \textit{offline} experiences of humans, the social behaviour and movement of people in the \textit{real world}. Online social networking sites facilitated easy collection of data and social graphs which motivated a large number of studies: from analyzing the structure of the networks, identifying the most influential people in a network, predictive models for inferring social connections to evolution studies for communities in social graphs. 

However, the properties of the online social networks may not necessarily apply to the real world. Even the social ties projected in the online world may not necessarily capture the social ties that exist in reality. Online Social Networks are curated and rather embellished versions of real-life mobility and social interactions. It is therefore imperative to bridge the gap between the amounts of research on online social networks and the real lives. This brings forth the need to examine the real world social networks and individual mobility patterns. Such \textit{real world mobility data} can be collected through various online services such as geo-tagged tweets, check-ins from Foursquare and Facebook or from mobile apps data such as whatsapp. Since this data is collected from the people who visit certain places at a certain time, the properties inferred from such data are applicable to the real world as opposed to the online social networks.

There are certain disadvantages of collecting data from various online websites. First, the data can be highly irregular depending on whether the user checked-in at particular places, whether the application was running at that time, whether the GPS was switched on by the user etc. Second, it does not track the fine-grained movements of the user at regular intervals. Third, often the check-in location is known however the duration spent at the location may remain unknown. Finally, the number of check-in locations visited by the user are unbounded and therefore it is relatively hard to attach a context to every location. 

We attempt to demonstrate the power of collecting \textit{fine-grained behavioral social network data from mobile phones of users}. We introduce \textbf{DAIICT Spatio-Temporal Network (DSSN)} - a spatiotemporal dataset which addresses the above challenges and produces fine-grained data with each location visited by the users  recorded at minute and regular intervals. The dataset is complimented by an extensive survey of the participants in which demographic, sociability and most importantly ground truth data about social ties with other participants is collected. We then turn to various experiments conducted on one of the possible lines of research with this dataset- finding the inter-relationship between reported friendship and the time spent together, introducing \textbf{Temporal Diversity} as a feature to infer the strength of friendship between two individuals and comparing its performance to measures of Diversity in previous works. We believe Temporal Diversity can be beneficial in environments where mobility is governed by schedules and time-tables much like in real life. We also explore the concept of Time Diversity and how it evolves with time. \textbf{Finally, we discuss the numerous different research avenues that can be pursued using such a fine-grained spatiotemporal dataset.}

\section{DAIICT Spatio-Temporal Network (DSSN) Dataset}
\subsection{Subject pool}
The subjects from this study consisted of students pursuing B-Tech ICT at DAIICT (Dhirubhai Ambani Institute of Information and Communication Technology), a university located in Gujarat, India. It has a residential campus which spans 60 acres and houses approximately 1,500 students. This choice of location provided several advantages. 

\begin{itemize}
\item The movements of the subjects were largely encapsulated within the geographical boundary of the campus. This provided us with a finite set of locations that are visited by the subjects at different times.
\item Each location inside the campus has an inherent function attached to it. For example, the library is used for reading purposes while the sports center is used for recreational purposes. 
\item The subjects of the study pursuing the B-Tech ICT program are compulsorily residing within the campus. The in-campus residence along with the distinct designated buildings for recreation, academics etc inside the campus make it an interesting and reliable microcosm of real world activities, movements and social behaviour.
\end{itemize}
The data was collected from 46 subjects between the months of March and May, 2016. For this paper's analyses, we used a subset of the data collected during the month of April, 2016. Out of the 46 subjects that participated in the study, 36 of them completed the survey conducted in July, 2016. The subjects volunteered to become part of the experiment.

\subsection{Important Statistics}
For each user, the timestamp, latitude and longitude, elevation, accuracy, satellites, network provider is recorded.
The total number of data points collected are 7,33,403. The total number of data points within the month of April are: 6,59,268 (this subset of data is used for the following analysis). The total number of subjects using the application to record data are 46. The data recorded varies in accuracy with an average accuracy of 36.0 meters.

\subsection{GPSLogger Software for Data Collection}
The data was collected from the Android-based mobile phones of the subjects. The subjects installed the GPS-Logger app which is available on playstore (\url{https://play.google.com/store/apps/details?id=com.crearo.gpslogger}). The application exploits the GPS capabilities of the mobile phones to log co-ordinates and runs as a background process at all times.

The following functionalities are present in the application:
\begin{itemize}
\item Logs data from the mobile phones of subjects at a regular interval of 2 minutes on local storage. Since the application is only programmed to log location data, the privacy invasion is minimum as compared to previous such approaches which log voice calls, messages, active applications, phone's charging status etc.\cite{nathaneagle} 
\item The data is sent to the server periodically (every 2 hours) if the mobile phone is connected to internet. If not, the data file is pushed to a queue and resent at a later time.
\item The subjects cannot switch off the application (without ``Force Kill" or un-installing the application). It shall restart in 30 minutes if the subject attempts to do so.
\item The battery consumption is optimized to last as long as possible.
\item It is a user-friendly application where the subject only has to press ``start logging" after installing the application. 

\end{itemize}
Since the application automatically restarts following any crashes, data losses mainly occur only due to powered-off devices. The application can be assumed to be running on the phone while the phone is powered-on, however, the accuracy of the dataset generated shall rely on the strength of the GPS signal captured. The strength of the signal shall vary based on the location as well as hardware of the phone. We therefore have to deal with the accuracy loss and random chunks of missing data that is characteristic of any real-time GPS data collection.

An anonymized version of the dataset can be downloaded at: (\url{https://github.com/deshanadesai/Geospat}).

\subsection{Ground Truth: Self Reported Survey data}
We conducted an online survey for the subjects who participated in the DSSN data collection. The survey is detailed and focuses on questions to report strength of friendship and estimated average proximity with each subject. It also includes general questions regarding the subjects' social behaviour, participation in various activities, anxiety levels, academic performance etc.
\newline\newline
\underline{Questions to be answered for each subject in the study.}
\begin{itemize}
\item Estimate your average proximity with the Person. (Time spent together on an average per day)
\begin{itemize}
\item Scale of 1-5 (1- 0 to 5 minutes, 2- 5 to 30 minutes, 3- 30 minutes to 2 hours, 4- 2 to 4 hours, 5- 4 hours and above)
\end{itemize}
\item Estimate strength of friendship with the person.
\begin{itemize}
\item Scale of 0-5 (0- Do not know the person, 1- Acquainted, 2- Sort of friends, 3- Friends, 4- Good friends, 5- Very good friends)
\end{itemize}
\end{itemize}

\underline{General Questions }
\begin{itemize}
\item Rate your participation in the following activities:
\begin{itemize}
\item Sports, Programming, Quizzing, Debate, Electronics, Writing college magazine, Dance, E-sports, Music, Drama, Academics, Research
\end{itemize}
\item Native Language, Birthplace
\item Rate amount of stress experienced , GPA (academic performance), productivity with the time in college,  satisfaction with time in college, social comfort, self-confidence.
\end{itemize}

\section{Problem Definition}
\subsection{Problem Statement}
Given a set of users U = (u1,u2,..,un) and a set of data points recorded
by every user u: Users location - \textit{l}, which consists of latitude and longitude values, the timestamp - \textit{t}, the provider used to measure location - \textit{p} and finally, the accuracy logged by the provider - \textit{A}.\newline\newline
\textbf{The objective is to infer the relationship score between each pair
of users based on quantitative values.} \newline\newline
\textit{Definition 1: } Relationship strength is a quantitative measure that tells
how strongly associated two people are. \newline\newline
\textit{Definition 2: } Encounter is defined as an event when two users co-occur
at the same place at the same time. The distance threshold \textit{d} chosen can be varied with the mean accuracy of the logged data points.

\subsection{Related Works}

One of the pioneering papers that study the behavioral characteristics of friendship and infer social network structure of the real world using Mobile Data is by Nathan Eagle et al\cite{nathaneagle}. They study social networks with binary relational ties (i.e. are two students friends or not?). However, these binary indicators only provide a very coarse indication of the nature of the relationships, and do not embrace the complexity of human relationships. Our purpose is to estimate the strength of people's relationships based on their interaction frequency, other proximity variables and describe the same in a discrete manner tending to the degree of friendship. Their collection of data includes mobile phone logs, calls, messages, usage of applications etc which is considerably privacy cannot be done at scale. Their results showed that the behavioural data collected by the mobile phones and the self-reported data are indeed related. In addition, the amount of communication was the most significant predictor of friendship.

This is extended by the work of Crawnshaw et al. who introduces various features such as specificity, location entropy, etc to analyze social connections. This study provides an insight into the social network structure showing that there exists a relationship between the mobility patterns of the user and number of friends that the user has in his social network. Further, Cyrus et al. \cite{EBM} included the impact of co-incidences and co-occurrences at locations to infer a continuous variable predicting the relationship strength between a pair of subjects. They compute the strength of relationship by conducting multiple linear regression over location diversity (the spread of encounters over different locations) and frequency of encounters weighted by the location entropy (how crowded a place is). 

\cite{cho2011friendship} explore the inherent strucutre in mobility patterns which are governed by geographic and social constraints. They find that short ranged travel is periodic both spatially and temporally and not affected by social ties, while long-distance travel is more influenced by social networkt ties. 

\subsection{Preliminaries}

\textbf{Temporal Representation: } The day is divided into intervals of \textit{t} minutes. For example, if \textit{t} is 5 minutes, the day can be divided into 288 intervals from 00:00-00:05, 00:05-00:10 to 23:55-00:00. \newline\newline
The variable \textit{t} can be a maximum of 1440 minutes where the entire day is counted as one interval. 
\newline\newline

\textbf{Temporal Encounter Vector: }Temporal Encounter Vector T$_{ij}$ contains the total number of times user i and user j had an encounter for each \textit{t} minute interval of the day over a period of N days. The day can be divided into 1440/\textit{t}  intervals and hence the dimensions of the vector are fixed. \newline\newline
Example, for t=5 minutes, 
T$_{(1,2)}$ = (1,3,0,0,0,0,...2) would mean that user 1 and user 2 had one encounter between 00:00-00:05, three between 00:05-00:10 and two between 23:55-00:00 across N days.
\newline\newline

\textbf{Temporal Diversity: }
We introduce a new feature called Temporal Diversity which can be beneficial in predicting social ties based on spatio-temporal data. 
\newline\newline
\textit{Motivation: } Over a period of days, encounters with a close friend will be spread across different time intervals in a day, as compared to encounters with someone you only meet because of scheduled activities. Essentially, encounters with friends over time don't follow any rules or schedules and are randomly spread across a day's time-span. On the other hand, encounters with people whom a person meets due to scheduled activities (eg: for work, lectures, tutorials) are routine and therefore would occur repeatedly at a scheduled time. We aim to capture this diversity in encounters over time to infer relationship strength.
\newline\newline
\textit{Definition: Temporal Diversity quantifies the effective spread of encounters across time-intervals in a day's timespan. }\newline

Given two users i and j,  Temporal Encounter Vector $T_{ij}$ contains the total number of times user i and user j had an encounter for each \textit{t} minute interval of the day over a period of N days.

Let \begin{equation}r_{i,j,l,t} = <i,j,l,t>\end{equation} represent an encounter between user i and user j in location l and time interval  \textit{t}. Let \begin{equation}R_{i,j}  = \bigcup_{i=1}^nr(i,j,l,t)\end{equation} be the set of co-occurrences of User i and j in all time intervals.

The probability that a randomly picked encounter from the set R(i,j) happened at time interval  t is:
\begin{equation}P_{(i,j,t)} = |R_{(i,j,t)}|/|R_{(i,j)}|\end{equation}

If we randomly pick an encounter from the set R(i,j) and define its time-interval as a random variable then the uncertainty associated with this random variable is defined by the Shannon entropy for user i and j as follows:
\begin{equation}H_{(i,j)}^{S} = - \sum_{t}(P_{(i,j,t)}logP_{(i,j,t)})\end{equation}

\begin{equation}D = exp(H)\end{equation}
Diversity \textit{D} is the effective number of t min time intervals user i and j have been together for in a day.

The more spread the encounters are across different time intervals, the higher the diversity.

Example: If we use the temporal representation with t=120 minutes, you can divide the day in 12 parts as follows: 00:00-02:00, 02:00-04:00 till 22:00-00:00.
\newline\newline
User A and User B have a Temporal Encounter Vector T$_{ab}$ calculated over 14 days such that, 
\newline\newline
T$_{ab}$ = (0,0,0,2,10,3,0,0,0,2,0,0) \newline
T$_{ac}$ = (3,0,4,0,3,0,2,2,0,0,0,3)
\newline\newline
Expanding T$_{ab}$ for clarity, User A and User B, had 2 encounters in 06:00-08:00 interval, 10 encounters in 08:00-10:00 interval, 3 encounters in 10:00-12:00 interval and 2 encounters in 18:00-20:00 interval  over 14 days. 
Note: User A and User B  have had 17 encounters in 14 days, so have User A and User C.
\newline\newline
Temporal Entropy for T$_{ab}$ = 1.1218 \newline
Temporal Diversity for T$_{ab}$ = 3.1
\newline\newline
Temporal Entropy for T$_{ac}$ = 1.7623 \newline
Temporal Diversity for T$_{ac}$ = 5.8 
\newline\newline
Most encounters for user T$_{ab}$ happened in the time interval 08:00-10:00 which could suggest that they were governed by some schedule such as a common lecture or breakfast routine. Encounters between User A and C are spread across the day and are more random or one might say \textit{diverse.}
\newline\newline
We want to give less weight to to time intervals where the users have encountered a lot because it could be due to some schedule shared between the users. They will still contribute, but lesser compared to encounters which were not part of schedules. Our assumption is that interactions with friends overtime are outside the purview of schedules and meeting times will be distributed across the entire day if averaged over N days.\\
\\
\\
\\
\\

\begin{figure}
\centering
\includegraphics[width=0.5\textwidth]{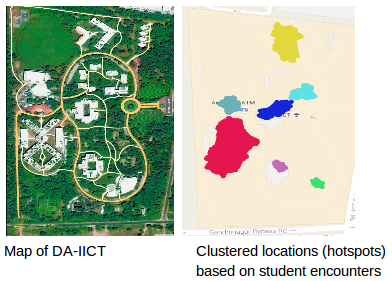}
\end{figure}

\section{Experiments with Temporal Diversity}
\subsection{Data Pre-processing}
We use the the DAIICT Spatio-Temporal Social Network (DSSN) with the following pre-processing:

\begin{enumerate}
\item All timestamps are rounded off to the nearest 5 minute interval to align data collected from different devices/users. 
\item The duplicate points occurring for a particular ID at any timestamp are dropped.
\item Only data points collected between 2016-04-01 00:00:00 and 2016-05-01 00:00:00 are used.
\item We keep the GPS points whose reported accuracy from the phone is less than 60 meters.
\item For a user, if he/she has less than 20\% data points collected for a day, we discard that day's data for the user. We assume the app malfunctioned or wasn't switched on for that day.
\item Users with data for less than 5 days are dropped from the analysis 
\item Only those Users are picked for which ground truth data is also available. 
\item Reported closeness \textit{Friends} and \textit{Sort of Friends} are clubbed into one category; leaving a total of 5 categories. 
\begin{table}
\centering
\caption{Re-grouping}
\begin{tabular}{|c|c|l|} \hline
\ Closeness & Description\\ \hline
\ 0 & Don't know the person\\ \hline
\ 1 & Acquaintance \\ \hline
\ 2 & Friends \\ \hline
\ 3 & Good Friends \\ \hline
\ 4 & Very Good Friends \\ \hline
\hline\end{tabular}
\end{table}
\end{enumerate}

It's not necessary that Users have data for all the 30 days for which the experiment was conducted. A fraction of users left/joined the experiment in the month of data collection. Hence, it's important to keep the number of common days between two people as part of the discussion as compared to the number of days a User had data overall. 
We choose N, the number of common days as a minimum 7. Selecting a number too less leaves us with less number of data points per pair and selecting a number too large leaves us with less number of User pairs. Hence, all our results reflect a scenario where any two Users have at least 7 days worth of data common between themselves.

It's important to note that reported closeness between two users is bi-directional. It's not necessary that User A and User B have reported closeness to each other equally. Hence, U$_{ab}$ and U$_{ba}$ are treated differently in the experiments.

\subsection{Evaluation}

We attempt to evaluate the effectiveness of Location Diversity \cite{EBM}, Temporal Diversity and Average Encounters per day as features in predicting reported closeness scores between Users. We primarily use F-Tests and pearson correlations.  

\subsubsection{Sweeping over width of time interval (t)}

The width of the time-intervals in which one divides the day in the temporal representation is an important parameter. For example, take User A and User B as discussed in the example with the Temporal Encounter Vector as T$_{ab}$ = (0,0,0,2,10,3,0,0,0,2,0,0). User A and User B have 10 encounters in 08:00-10:00 interval. We assume User A and User B have a common lecture at this time.  If the interval width was shorter such as 5 minutes, there is a likelihood that these 10 encounters would be distributed across the intervals 08:00-08:05, 08:05-:08-10 to 09:55 to 10:00. Both User A and User B may not enter the classroom at the same time or may not always be in each others exact vicinity for an encounter to be recorded. So, even though the meeting is a scheduled encounter, the Temporal Diversity is not able to account for it, rather in all likelihood Temporal Diversity increases positively because of it.

As we saw, this effect can be mitigated by having wider time-intervals. But if the time-intervals are kept too wide, the actual random encounters across the day would be grouped into one and information would be lost.

We calculate for  Temporal Diversity scores for the same set of encounters at different widths \textit{t}. The cross correlation between each regressor and the target is computed, which is eventually converted to a F score and a p-value.

\begin{table}
\centering
\caption{Sweeping over width of time interval (t)}
\begin{tabular}{|c|c|l|} \hline
\ Width t (minutes) & F Value & p-value \\ \hline
\ 5 & 53.44 & 0 \\ \hline
\ 15 & 63.47 & 0 \\ \hline
\ 30 & 67.45 & 0 \\ \hline
\ 60 & 69.83 & 0 \\ \hline
\ 90 & 64.97 & 0 \\ \hline
\ 120 & 63.00 & 0 \\ \hline
\ 180 & 54.04 & 0 \\ \hline
\ 240 & 48.15 & 0 \\ \hline
\ 360 & 54.05 & 0 \\ \hline
\ 710 & 37.09 & 0 \\ \hline
\hline\end{tabular}
\end{table}

\textbf{Observation: } We notice, 60 minutes has the highest F-value and hence can be the optimum width t for our dataset. F-values increase from 5-minute interval to 60-minute interval and thereafter decrease till the 720 minute interval.

\subsubsection{Predicting Reported Closeness through Regression.
}

\textbf{Location Diversity: } We calculate Location Diversity based on the explanation given in EBM\cite{EBM}. The only difference being we used geohashing\cite{wikigeohash} to divide the campus into fixed regions to map each encounter to a location id. We use hashes of length 8 which give a precision of 19 meters more or less.

We apply F-test on Location Diversity, Temporal Diversity and Mean Encounters each regressing them with the outcome (closeness).

\begin{table}
\centering
\caption{Predicting Reported Closeness through Regression}
\begin{tabular}{|c|c|c|l|} \hline
\ Feature & F Value & p-value & Correlation (r) \\ \hline
\ Location Diversity & 42.01 & 0 & 0.23 \\ \hline
\ Mean Encounters & 35.10 & 0 & 0.21 \\ \hline
\ Temporal Diversity (t=60) & 69.83 & 0 & 0.30 \\ \hline
\hline\end{tabular}
\end{table}

\textbf{Observations: } We notice that amongst the three features, Temporal Diversity is ranked the highest, followed by Location Encounters and Mean Encounters.  Temporal Diversity is most correlated with closeness. \\
\\

We further explore how Temporal Diversity, Location Diversity and Mean Encounter scores are distributed across different closeness sub-groups.

\begin{figure}
\centering
\includegraphics[width=0.5\textwidth]{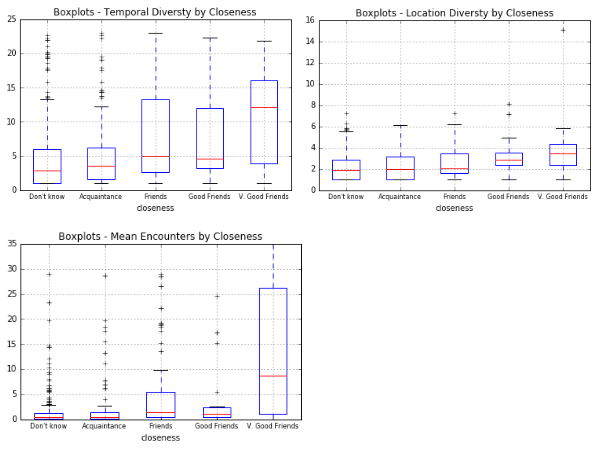}
\caption{Diversities, Mean Encounters versus Reported Closeness}
\end{figure}

\textbf{Observations: }
\begin{itemize}
\item Mean Temporal Diversity for sub-group \textit{Very Good Friends}, is significantly higher than the rest. 
\item There are many outliers in \textit{Don't Know} and \textit{Acquaintance} sub-group for Temporal Diversity. It suggests there are a lot of coincidences happening. It will be worthwhile to explore further ways of accounting for co-incidences in encounters. Renyi Entropy has been explored previously\cite{EBM} to tackle for co-incidental encounters but in domain of locations. We apply the same concept to the domain of time.
\item Temporal Diversity isn't able to differentiate between \textit{Friends} and \textit{Good Friends}. The overlap between both is high. 
\item Location Diversity shows a healthy increase in mean diversity across sub-groups but contains a lot of overlaps between different sub-groups both in the lower quartile and the upper quartile.
\item Mean Encounters for \textit{Very Good Friends} is significantly different than the other sub-groups. It would suggest that pairs in that sub-group had a healthy amount of encounters every common day.  Strikingly, the same effect is not visible for pairs in  \textit{Good Friends} or even \textit{Friends}, where the median of Mean Encounter is approximately 1 as compared 8 for \textit{Very Good Friends}. But the same sparsity in encounter data doesn't negatively affect temporal diversity for the said sub-subgroups. We can say Temporal Diversity is robust to sparsity in the daily data which is an important factor in real life geospatial applications.

\end{itemize}

\subsubsection{Renyi Entropy and Co-incidences} 
We use Renyi Entropy as used in Cyrus et. al\cite{EBM}, which is a generalization of Shannon entropy.
Let R$_{ij}$ be the diversity calculated from the Renyi entropy.

\begin{equation}H_{ij}^R =  [-log \sum_{t}{P_{(i,j)}^t}^q ]/(q-1))\end{equation}
\begin{equation}R_{ij} = exp(H_{ij}^R)\end{equation}
\begin{equation}R_{ij} = exp[(-log\sum_{t}{P_{ij}^t}^q)/(q-1)]\end{equation}
\begin{equation}D_{ij} = [\sum_{t}((P_{ij}^t)^q)^(1/(1-q))]\end{equation}

Cases of Renyi Entropy:
\begin{enumerate}
\item As q approaches zero, the Renyi entropy increasingly weighs all possible events more equally, regardless of their probabilities. In the limit for q -> 0, the Renyi entropy is just the logarithm of the size of the support of X. 
\item When q < 1, the temporal diversity tends to give more weight
to the local frequencies with low-values in the Temporal Encounter vector. In other words, the lower the number of times a pair of students have met in a particular time slot, the more weight it gets from the diversity or the more impact the local frequency can make on diversity.
\item The limit for q -> 1 is the Shannon entropy. 
\item When q > 1 the Renyi entropy H$_{ij}$ , the opposite of q < 1 occurs and consequently the diversity D$_{ij}$ , more favorably considers the high values of the Temporal Encounter vector. 
\item As q approaches infinity, the Renyi entropy is increasingly determined by the events of highest probability.
\end{enumerate}
When q<1 is used, Renyi Entropy and consequently the diversity, more favourably considers lower values of  local temporal frequencies than higher values. Using q = 0.5

\cite{EBM} goes in detail about its exploration on how the effect of co-incidences can be controlled by using Renyi entropy based Diversity in the domain of locations. We apply to same concept for Temporal Diversity and sweep over the the parameter \textit{q}, the order of diversity, which decides the senstivity of the final Temporal Diversity to the number of encounters in each time-interval.   

Refer to Table \ref{renyi} for results.

\textbf{We observe that the lower the order of the diversity (\textit{q}), the more effective is Temporal Diversity in predicting closeness.} This is in line with the results noticed in \cite{EBM} for Location Diversity. 

\begin{table}
\centering
\caption{Sweeping over order of diversity (d)}
\label{renyi}
\begin{tabular}{|l|l|}
\hline
Order of Diversity & F-Value \\ \hline
0.1                & 79.82   \\ \hline
0.2                & 78.18   \\ \hline
0.3                & 76.66   \\ \hline
0.4                & 75.26   \\ \hline
0.5                & 73.98   \\ \hline
0.6                & 72.83   \\ \hline
0.7                & 71.78   \\ \hline
0.8                & 70.84   \\ \hline
0.9                & 69.99   \\ \hline
1.1                & 68.51   \\ \hline
1.2                & 67.87   \\ \hline
1.3                & 67.29   \\ \hline
1.4                & 66.76   \\ \hline
1.5                & 66.27   \\ \hline
1.6                & 65.82   \\ \hline
1.7                & 65.41   \\ \hline
1.8                & 65.02   \\ \hline
1.9                & 64.66   \\ \hline
2                  & 64.33   \\ \hline
\end{tabular}
\end{table}

\subsubsection{Changing Temporal Diversities Over Time.}
How does Temporal Diversity between two people change over time? We've emphasized before how the essence of Temporal Diversity lies in the growing randomness of encounters across a day over time for users with strong social ties. Hence, it's important to explore how temporal diversity changes for over days for different sub-groups of closeness. 

\textbf{Methodology:} Each user has different number of common days for which they share data. For this experiment, we calculate temporal diversity between pairs of users using only common days less than equal to d, while we sweep d from 1 to 11.  
For example, for \textit{d}=3 we calculate temporal diversity between pairs based on data from their respective first three common days. In case a pair has only 2 common days, temporal diversity for \textit{d}=3 is set as empty. Then, for each sub-group in closeness and common day  \textit{d}, we calculate the average temporal diversity.

\begin{figure}
\centering
\includegraphics[width=0.5\textwidth]{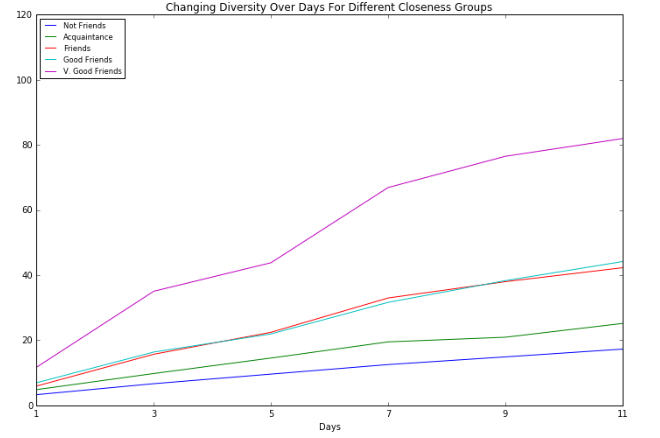}
\caption{Changing Diversities over Days for different Groups}
\end{figure}

\textbf{Observations: }
\begin{itemize}
\item Temporal Diversity in general increases as the number of common days between two users increases.
\item For the sub-group \textit{Very Good Friends}, average temporal diversity is starkly higher than the averages in other sub-groups. 
\item Average temporal diversity of encounters between \textit{Very Good Friends} increases at a much higher rate than that of other sub-groups.
\end{itemize}

\section{Conclusions}
In this paper, we first introduce and release the DAIICT Spatio-temporal Social Network (DSSN) dataset which is a granular dataset about 46 participant's movement in a residential college campus over a span of thirty days.  It reflects human patterns guided by routine and schedule in a geographically bounded area of a university. The data is complemented with rich self reported data about the participants based on an extensive survey capturing demographics, extent of participating in various campus activities, proximity to other participants in the survey and questions pertaining to happiness, sociability and self-confidence.

Further, we introduce a new feature called Temporal Diversity which our experiments confirmed to have superior predictive power for inferring social ties as compared to its counterparts: Location Diversity and Mean Encounters. Next, we examined how Temporal Diversity changes with time and gets better at differentiating between sub-groups of social-ties as the number of common days increase. Lastly, we saw how Temporal Diversity is robust to sparsity of data collected between two users and can function well even with lesser data. We believe, Temporal Diversity can be an useful feature to predict social ties in environments where users are governed by fixed schedules and the number of meaningful unique locations are less. We believe a bustling resendential university campus is a microcosm of urban lifestyles to some extent.

This work along with releasing of the DSSN dataset opens up opportunities to answer interesting multi-disciplinary questions. How does mobility affect self-confidence, sociability, happiness or GPA? Do people with similar interests end up in the same social groups? Do people in relationships have a particular mobility pattern? Are certain time-intervals of meeting more helpful in predicting social ties than others? How can one effectively predict social-ties in an environment where co-incidental encounters are abundant due to schedules and closed spaces like in workplaces? We wish to investigate some of these issues in our future work and at the same time hope that the research community finds the dataset useful to explore some of their own questions.


\section{Acknowledgments}
The authors would like to heartily thank  professor Sanjay Srivastava of DA-IICT for his encouragement, initiatives and help. We also thank the software creators of the publicly available GPS Logger (\url{https://github.com/mendhak/gpslogger}) and Vraj Delhivala for the additions to the software application. Finally, we are thankful to all the participants for there help with the data collection and survey.


\medskip
 
\begin{thebibliography}{9}
\bibitem{crandall:coincidences} 
D. Crandall, L. Backstrom, D. Cosley, S. Suri, D. Huttenlocher and J. Kleinberg- \textit{Inferring social ties from geographic coincidences}. 
Proc. National Academy of Sciences,December, 2010.

\bibitem{nathaneagle} 
Nathan Eagle , Alex (Sandy) Pentland , David Lazer\textit{Inferring Social Network Structure using Mobile Phone Data}.

\bibitem{hawkes} 
Charles Blundell, Katherine A. Heller, Jeffrey M. Beck- \textit{Modelling Reciprocating Relationships with Hawkes Processes}.

\bibitem{lsh} 
Zhenyu Wu , Ming Zou- \textit{An incremental community detection method for social tagging systems using locality-sensitive hashing}.

\bibitem{brugere} 
Ivan Brugere, Venkata M. V. Gunturi, Shashi Shekhar- \textit{Modeling and analysis of spatio-temporal social networks}.

\bibitem{EBM} 
Huy Pham, Cyrus Shahabi, Yan Liu- \textit{EBM - An Entropy-Based Model to Infer Social Strength from Spatiotemporal Data}.

\bibitem{summary} 
Huy Pham, Ling Hu, Cyrus Shahabi- \textit{Towards Integrating Real-World Spatiotemporal Data with Social Networks}.

\bibitem{smallgroups} 
Nancy Katz, David Lazer, Holly Arrow, Noshir Contractor- \textit{Network Theory and Small Groups}.


\bibitem{usersimilarity} 
Quannan Li, Yu Zheng, Xing Xie,Yukun Chen, Wenyu Liu, Wei-Ying Ma- \textit{Mining User Similarity Based on Location History}.

\bibitem{hecking} 
Tobias Hecking, Tilman Göhnert, Sam Zeini, Ulrich Hoppe- \textit{Task and Time Aware Community Detection in Dynamically Evolving Social Networks}.

\bibitem{wikigeohash}
\textit{Wikipedia - Geohash --- Wikipedia, The Free Encyclopedia, [Online; accessed 20-August-2016]}.
 
\end{thebibliography}
\end{document}